# Research on the recommendation framework of foreign enterprises from the perspective of multidimensional proximity

**Liang, Guoqiang** [1], **Xie, Jiarui** [1], **Li, Mengxuan** [2] **and Zhang, Shuo** [1,*]

[1] Beijing University of Technology; Lianggq@bjut.edu.cn; 22117127@emails.bjut.edu.cn;

[2] Langfang Normal University; 3454261499@qq.com;

* Zhang, S; zhangshuozs@bjut.edu.cn; Tel.: +86 18210125514

**Abstract:** As global economic integration progresses, foreign-funded enterprises play an increasingly crucial role in fostering local economic growth and enhancing industrial development. However, there are not many researches to deal with this aspect in recent years. This study utilizes the multidimensional proximity theory to thoroughly examine the criteria for selecting high-quality foreign-funded companies that are likely to invest in and establish factories in accordance with local conditions during the investment attraction process.First, this study leverages databases such as Wind and Osiris, along with government policy documents, to investigate foreign-funded enterprises and establish a high-quality database. Second, using a two-step method, enterprises aligned with local industrial strategies are identified. Third, a detailed analysis is conducted on key metrics, including industry revenue, concentration (measured by the Herfindahl-Hirschman Index), and geographical distance (calculated using the Haversine formula). Finally, a multi-criteria decision analysis ranks the top five companies as the most suitable candidates for local investment, with the methodology validated through a case study in a district of Beijing.The example results show that the established framework helps local governments identify high-quality foreign-funded enterprises.

## 1. Introduction

Attracting investment has long been a key issue of concern for enterprises and governments worldwide. Marx argued that while profit rates tend to equalize across different regions over time, there nevertheless exist temporal and spatial opportunities for higher profit rates[1]. In this context, various scholars have proposed their own theories, each emphasizing the importance of investment attraction from different perspectives. For instance, Professor Kiyoshi Kojima of Hitotsubashi University suggested that international direct investment should first be directed towards industries in the home country (the investing country) that are already in or are about to enter a relatively disadvantaged position (i.e., marginal industries). These marginal industries, in turn, are also those in which the host country holds or has the potential for comparative advantage[2]. Furthermore, Professor Raymond Vernon of Harvard University, building upon the theory of monopolistic advantages, proposed the Product Life Cycle Theory. According to Vernon, attracting foreign investment is aimed at facilitating the combination of capital and technology with the local workforce, which is often available at lower costs[3].

The theory of multidimensional proximity is an important theoretical basis for analyzing the economic development of different regions. Kirat and Lung argue that multidimensional proximity consists of three main forms: geographical proximity, organizational proximity, and cognitive proximity[4]. Boschma, on the other hand, suggests that the concept of proximity can be distinguished into geographical, organizational, cognitive, social, and institutional proximity[5], which is also the primary theoretical framework applied in this article. Regarding geographical proximity, Torre (2005) defines it as "the kilometer distance between two units" [6], while Yang Guibin and Li Wanhong regard geographical proximity as an endogenous variable used to describe the economic relations between regions [7].

However, regardless of the specific definition, the academic consensus is that the closer the spatial distance between locations, the more beneficial the economic exchange and development between regions. As for organizational proximity, the most detailed definition currently available is provided by Torre, who defines it as the presence of similar rules and systems between entities [6]. Organizations with similar structures are more likely to engage in cooperative exchanges. In the case of institutional proximity, its dual analysis includes organizational rules and conventions, which, to some extent, can be viewed as a form of organizational proximity [8]. Finally, cognitive proximity primarily focuses on the discussion of people. Wuyts defines cognitive proximity as "the similarity in the way entities perceive, explain, understand, and assess the world" [9].

In existing research, the impact of different types of proximity on investment attraction varies. Geographical proximity is considered to be particularly helpful in fostering financing relationships. For instance, Eva Lutz and others argue that in the venture capital sector in Germany, geographical proximity increases the likelihood of investments being made to the target companies[10]. Similarly, Yongzhe Yan and colleagues suggest that, within the context of technological innovation, geographical proximity between technology firms significantly influences technological innovation [11]. In contrast, organizational proximity typically mitigates the negative impacts of geographic and cultural distances in cross-border investments through trust established between firms and even governments. This trust in institutional organizations is especially important for investments in emerging economies, while relational trust tends to be more significant in developed economies [12]. Such trust can facilitate smoother cross-border interactions and collaborations. As for social proximity, it is related to the sharing of knowledge and understanding, which in turn enhances collaborative innovation performance between organizations. However, its effectiveness may be influenced by the inefficiency of internal networks within organizations [13].

At the same time, different types of proximity also interact with each other, exhibiting complementary effects. For example, Stefano Usai and colleagues, in their study of knowledge flows between company agreements, explored the positional relationships of different proximities within corporate networks. They found that the five dimensions of proximity jointly play a positive and interrelated role in determining the likelihood of knowledge exchange between companies, suggesting that these proximities complement each other rather than serve as alternative channels [14].

Therefore, this paper argues that the interaction of multidimensional proximity plays a facilitative role in promoting cooperation and communication between regions. Based on this, the paper aims to provide a simple and feasible practical approach for attracting investment. Accordingly, the structure of the remaining sections is as follows: Section III introduces the research framework; Section IV focuses on the construction of Chaoyang District's foreign investment attraction network as the research subject. It develops a framework for selecting the key nodes in Chaoyang's investment attraction network using detailed quantitative indicators to verify the framework's feasibility; Section V concludes the paper.

## 2. Materials and Methods

### *2.1. Data source*

This study's data consists of two main components: the Wind Financial Terminal database and the Osiris database. The first component includes enterprise data from the district, focusing on companies currently registered in the district's archives and their national standard industry classifications. This study uses Chaoyang District of Beijing, China, as the example. The selection criteria were as follows: 1) Companies that were operational in 2022, and 2) Companies with a unified social credit code. Based on these criteria, a total of 4,035 enterprises in the district were identified. The second component encompasses data on the world's top 5,000 companies, detailing their registered addresses, profit status, international industry classifications, and current market values. The selection criteria included: 1. Companies that are operational (72,913), and 2. The top quarter of companies (15,656) based on net revenue for 2022, measured in USD. This process resulted in the identification of 15,656 eligible listed companies.

### *2.2 Research Framework*

To create an investment network framework that aligns with the economic landscape and industry within the specific region, the significance of foreign-funded enterprises in regional industrial development will be illustrated through clear scoring. This paper distills and refines four key indicators from research on enterprise development: the compatibility of industry with regional industrial planning, the revenue scale of enterprises, the proximity of foreign-funded enterprises' registered addresses to China's strategic planning, and the strategic positioning of these enterprises. A two-stage method is employed to address the varying complexities of these indicators, facilitating the selection of foreign-funded enterprises. The resulting comprehensive filtering framework is depicted in Figure 1.

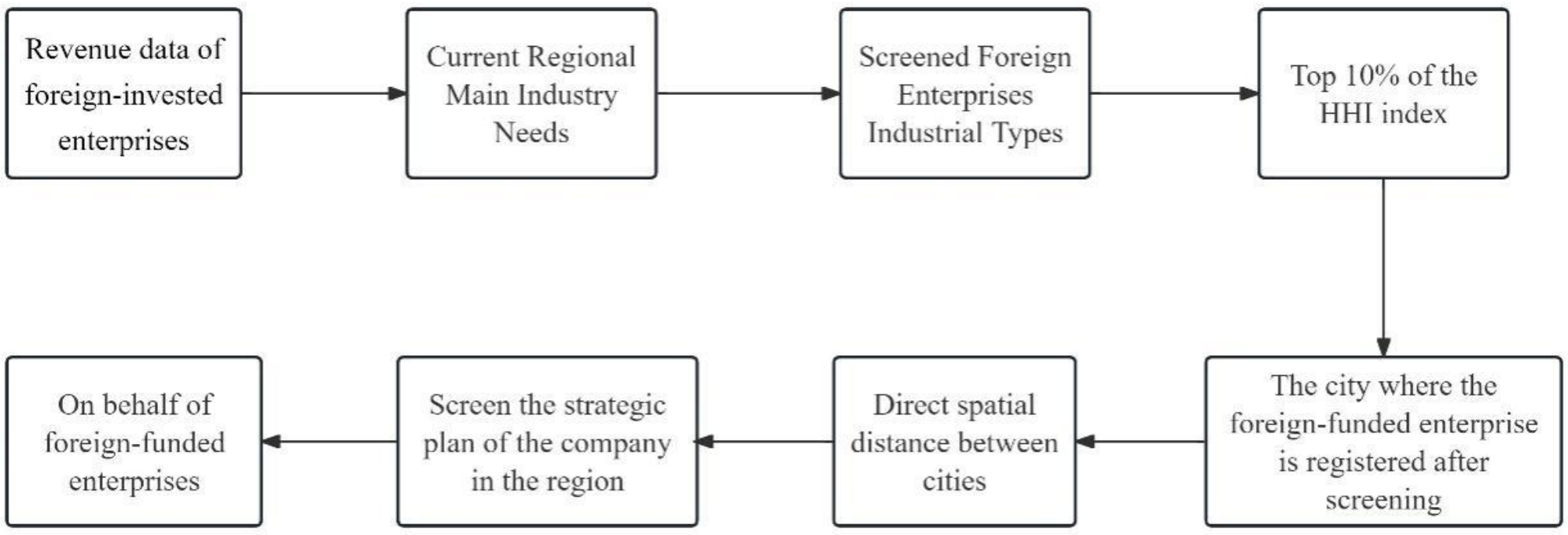

Figure 1. Flow chart of investment network framework construction in district

In the examination of the appropriateness of revenue scale and industry, this study integrates the concepts of organizational proximity and foreign capital enterprises. The findings indicate that a larger revenue scale correlates with a greater alignment between the core business of the enterprise and the regional demand industry. This alignment facilitates improved resource allocation, enhances adaptability, and increases the likelihood of selecting the regional government as an investment target. Regarding the assessment of spatial distance between the city of enterprise registration and the research area, Boschma posits that geographical proximity can be characterized by either absolute distance (e.g., miles) or relative distance (e.g., travel time)[5]. Scherngell and Barber employ spherical distance as a metric for geographical distance[15].

In this analysis, the distance measured by Scherngell and Barber is utilized to evaluate the spatial separation based on the kilometre distance between two cities. According to the theory of geographical proximity, a smaller spatial distance to the research area increases the likelihood of attracting investment within that area. Furthermore, the analysis of strategic planning among enterprises in China, in conjunction with the theory of institutional proximity, suggests that a code of conduct that is particularly suited to the Chinese context is preferentially adopted to establish a foundational level of trust and mitigate transaction costs[16].

### *2.2. Index System*

#### 2.2.1. Industry revenue

The Helduff-Hirschman Index (HHI)[17] is commonly utilized by economists and government regulatory agencies to assess the market share of foreign-funded companies and the concentration of industries on an international scale.

$$HHI = \sum_{i=0}^{N}\left(X_i/X\right)^2 = \sum_{i=0}^{N} S_i^2 \qquad (1)$$

In this context, X denotes the overall market size of the industry, while signifies the market size of the i company within that industry. The current market value reflects the market size of a specific enterprise, and $S_i = X_i/X$ represents the market share of the i company. The variable n indicates the total number of companies operating in the industry.

The HHI index is used not only to indicate the size of each industry but also to assess the global concentration within those industries, reflecting their level of monopoly and competitiveness. By comparing the HHI index values numerically, various recommendations are made based on the varying degrees of industrial concentration and the current development status of industries in the district.

2.2.2. Subsection

The national industry classification standard was utilized to categorize businesses in the district. This standard was jointly released by the AQSIQ and the National Standards Commission on June 30, 2017, in accordance with the current Classification of National Economic Industries (GB/T 4754-2017) [18]. Industries are generally organized into categories based on the nature of their economic activities, which are further divided into major, middle, and minor classes, with the major category being used for classifying enterprises in Chaoyang District. For foreign enterprises, the classification follows the US Standard Industry Classification (US SIC) primary code. The US SIC is an initiative aimed at creating new measurement standards for the North American Free Trade Agreement (NAFTA) to enhance data comparability with other trade agreements. SIC codes consist of four digits, where the first two digits typically indicate broad industry categories and the last two digits provide more detailed subcategories. The term ‘primary code’ usually refers to the first two digits, representing a wide array of industry categories.

2.2.3. Spatial distance between regions

The Haversine formula, which was first published in English by James Andrew in 1805, is commonly used to determine the spatial distance between two cities by calculating the great circle distance between two locations on the Earth's surface [19].

$$a = \sin^2\left(\frac{\Delta lat}{2}\right) + \cos(lat1) \times \cos(lat2) \times \sin^2\left(\frac{\Delta lon}{2}\right) \tag{2}$$

$$c = 2 \times atan2(\sqrt{a})\sqrt{1-a} \tag{3}$$

$$d = R \times c \tag{4}$$

In this context, $\Delta lat$ and $\Delta lon$ denote the variations in latitude and longitude, respectively, between the two cities. The symbol R signifies the average radius of Earth, which is approximately 6,371 km. The variable d represents the calculated distance.

## 3. Empirical study: a case from Chaoyang district

### *3.1. Empirical research*

In 2018, Chaoyang District conducted a comprehensive review and publicly disclosed its policies and achievements related to attracting foreign investment since 2015. This review emphasizes key themes such as investment attraction, expansion of the service industry, industrial restructuring, creation of a business-friendly and livable environment, international openness, and globalization. By 2023, Chaoyang District had established itself as the location for all foreign embassies in China, with the exception of Russia, as well as hosting 90% of foreign media outlets in Beijing and 80% of international organizations and chambers of commerce. The district witnessed the establishment of 2,064 new foreign-funded enterprises, the presence of 146 multinational corporate headquarters, and a total contracted foreign investment amounting to 127.51 billion USD, with actual foreign investment reaching 73.97 billion USD. To facilitate this growth, Chaoyang District has implemented and continuously developed its investment policies, transitioning from the "two zones" policy to a more refined approach.

The investment environment was optimized through a shift from the "9+N" policy framework to the "5+3+N" model. Additionally, the district has enhanced its infrastructure and support services, including international medical, educational, and business facilities, to attract more foreign enterprises and international talent, thereby establishing a robust foreign investment network. Considering the complexities and uncertainties of the post-pandemic era, it is imperative for the Chaoyang District government to adapt its investment policies and strategies in response to evolving domestic and international conditions.

### *3.2. Industry suitability*

By analyzing statistics on the number of businesses within the same category and creating a bar chart, we can initially determine the quantity and registered capital of enterprises in various sectors within the Chaoyang District, as illustrated in Figure 2. The findings indicate that the primary industries in Chaoyang District include manufacturing, scientific research and technical services, information transmission, software and IT services, construction, leasing services, and retail and wholesale industries. Gephi software was utilized to visualize the collaboration network among different industrial sectors in the Chaoyang District, as depicted in Figure 3. When examining both charts, two key characteristics of the industrial cooperation network in Chaoyang District emerged: 1. Traditional industries remain the dominant sectors, but exhibit fewer connections and weaker links in the cooperation network, suggesting a low level of collaboration among these enterprises. 2. There is a strong interconnection among science, technology, and technical service companies.

The cooperation network reveals that the scientific research and technology service sector, along with the science and technology extension and application service sector, collaborate closely, whereas the Internet, information transmission, and software development sectors maintain relatively stable partnerships. It is suggested that the demand for foreign-funded enterprises in Chaoyang District should be evaluated from two perspectives: 1. The need for foreign investment in traditional industries is less than that in the technology and information sectors, as there are many traditional industries in Chaoyang District with limited collaboration among them, resulting in a smaller impact of traditional foreign-funded enterprises on the district's economic growth compared to technology and information firms.

Given the distinct characteristics of these two industries, it is recommended to implement competition policies for traditional sectors and selectively introduce a few foreign-funded enterprises to enhance market dynamics. Conversely, for the science, technology, and information sectors, efforts should be made to seek foreign partnerships and exchange advanced practices.

Table 1: Number of national standard industries - categories of enterprises in Chaoyang District

| National standard industry category | Counting item: National standard industry – category |
|---|---|
| Scientific research and technical services | 2677 |
| Information transmission, software and information technology services | 479 |
| Leasing and business services | 206 |
| Culture, sports and entertainment | 195 |
| Manufacturing industry | 172 |
| Wholesale and retail trade | 114 |
| Building industry | 63 |
| Water, environment and utilities management | 34 |
| Electricity, heat, gas and water production and supply industry | 20 |
| Banking industry | 18 |
| Education | 15 |
| Health and social work | 10 |
| Transportation, warehousing and postal services | 9 |
| Mining industry | 8 |
| Residential services, repairs and other services | 5 |
| Real estate industry | 4 |
| Agriculture, forestry, animal husbandry and fishery | 4 |
| Accommodation and catering | 2 |

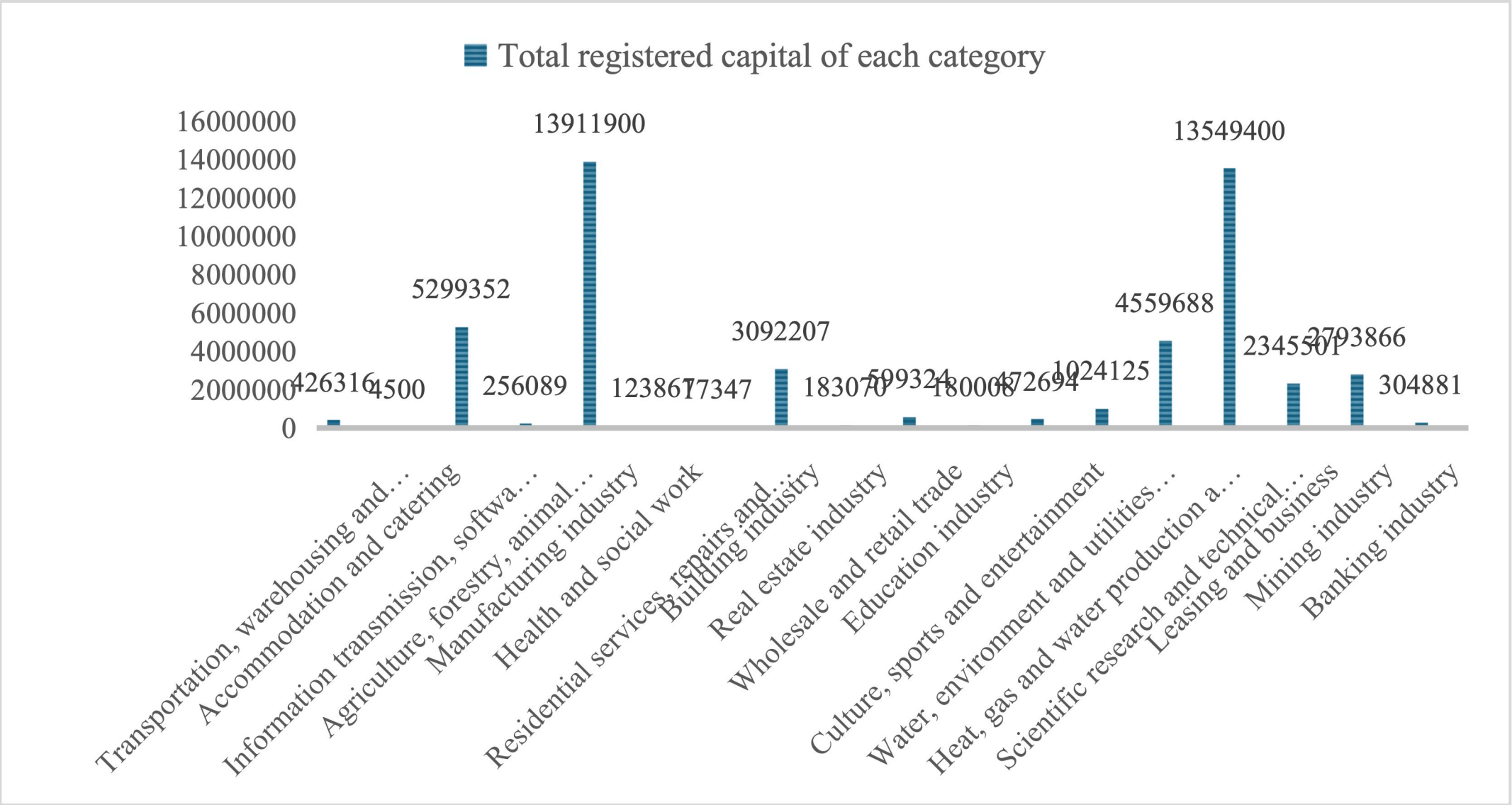

Figure 2: Total registered capital of national standard industries – categories in Chaoyang District

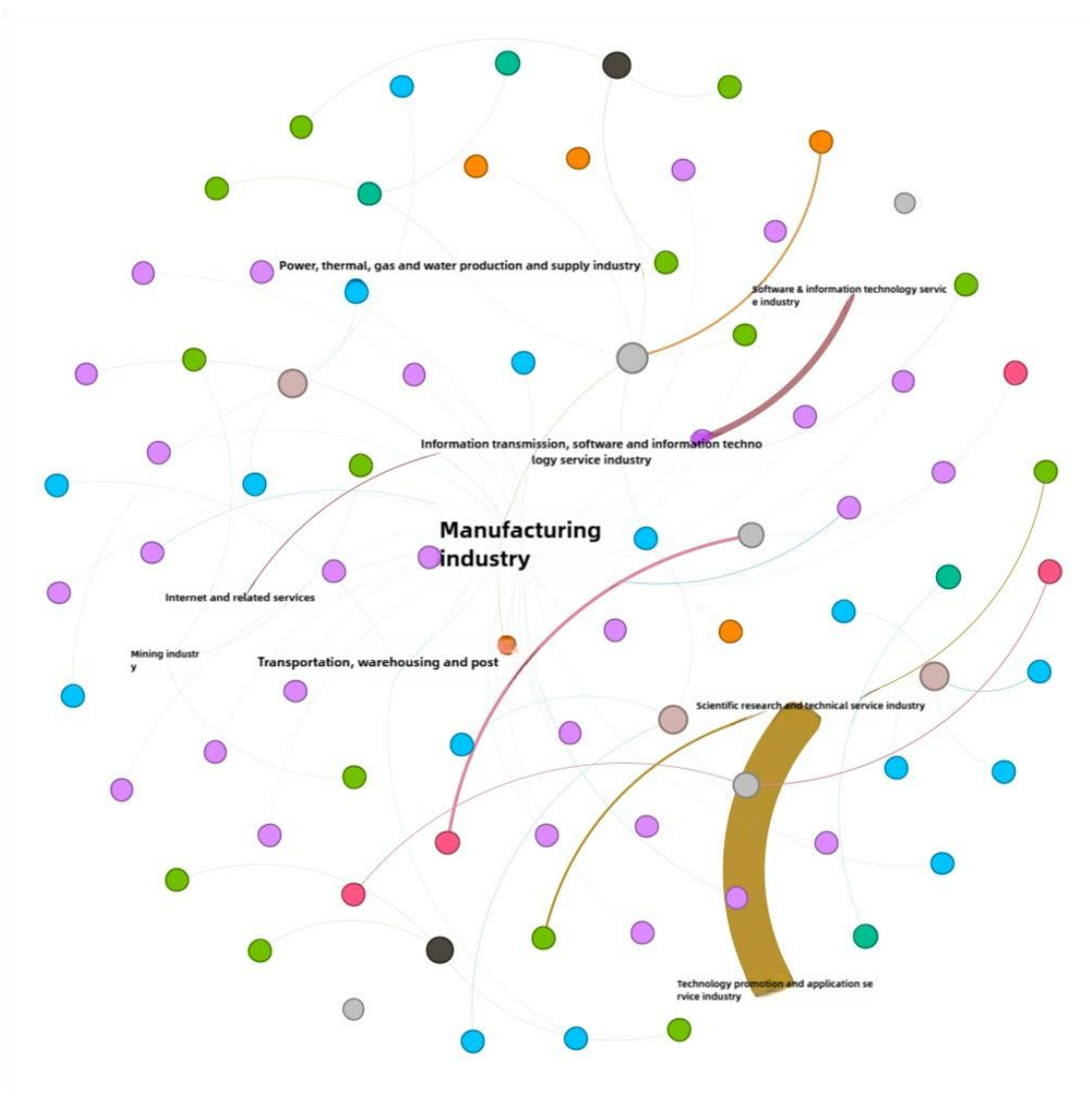

Figure 3: Chaoyang District industry national standard industry-category cooperation network

Owing to the varying representations of the GB industry and US Standard Industry Classification (US SIC), the SIC code is aligned with the GB industry. As per the USSIC primary industry code division standard, the initial two digits of the primary code indicate the industry classification, which is correlated, as illustrated in Table 2.

Table 2: The first two digits of the US SIC code match the national standard industry standard

| National standard industry – category | US SIC |
|---|---|
| Transportation, warehousing and postal services | 40, 41, 42, 43 |
| Accommodation and catering | 70, 58 |
| Information transmission, software and information technology services | 48, 73 |
| Agriculture, forestry, animal husbandry and fishery | 01, 02, 07, 08 |
| Manufacturing industry | 20, 21, 22, 23, 24, 25, 26, 27, 28, 29, 30, 31, 32, 33, 34, 35, 36, 37, 38, 39, |
| Health and social work | 80 |
| Residential services, repairs and other services | 72, 76 |
| Building industry | 15, 16, 17 |
| Real estate industry | 65 |
| Wholesale and retail trade | 50, 51, 52 |
| Education | 82 |
| Culture, sports and entertainment | 79 |
| Water, environment and utilities management | 49 |
| Electricity, heat, gas and water production and supply industry | 49 |
| Leasing and business services | 73 |
| Mining industry | 10, 12, 14 |
| Banking industry | 60 ,61, 62, 63, 64, 65, 67 |
| Scientific research and technical services | 87 |

All foreign-funded companies were categorized, and those with registered addresses in Chinese cities were omitted to determine the number of foreign-funded enterprises across different sectors. Owing to the significant amount of missing registered capital information for these companies, only the number of enterprises in each industry category is utilized to illustrate the international industrial distribution, as presented in Table 3.

Table 3: Number of foreign-funded enterprises by country

| **National standard industry – category** | **Number of enterprises** |
|---|---|
| Manufacturing industry | 4980 |
| Banking industry | 2938 |
| Information transmission, software and information technology services | 2027 |
| Mining industry | 1352 |
| Leasing and business services | 1034 |
| Scientific research and technical services | 865 |
| Wholesale and retail trade | 418 |
| Real estate industry | 388 |
| Building industry | 258 |
| Accommodation and catering | 233 |
| Water, environment and utilities management | 226 |
| Health and social work | 191 |
| Culture, sports and entertainment | 144 |
| Education | 62 |
| Transportation, warehousing and postal services | 59 |
| Residential services, repairs and other services | 43 |

The analysis indicates that the industrial distribution of foreign-funded enterprises largely mirrors that of Chaoyang District, with key sectors including manufacturing, mining, finance, leasing and business services, information transmission, software and IT services, and scientific research and technology services. However, there are notable differences in industrial demand within Chaoyang District.

Firstly, There is a higher presence of foreign-funded enterprises in the financial and mining sectors than in Chaoyang District. Given Beijing's environmental governance policies and Chaoyang District's efforts to develop an international double-first-class business district, it appears that the demand for foreign investment in the mining sector is lower than that in the financial sector. Therefore, focusing on financial industry enterprises could provide new momentum to the Chaoyang CBD.

Secondly, While the Chaoyang District has a limited number of manufacturing industries with relatively low registered capital, there are a significant number of foreign-funded enterprises in the international market with substantial registered capital. To address the industrial demand in Chaoyang District, it is essential to consider specific types of international manufacturing industry. Thus, attention should be paid to the current state of the manufacturing sector and the leading enterprise types in the international market, followed by further analysis.

In general, considering the key industries and the number of foreign-funded companies in different sectors within Chaoyang District, it is thought that foreign-funded enterprises in leasing and business services, information transmission, software and IT services, and scientific research and technology services demonstrate a notable level of adaptability. This adaptability should be evaluated alongside the market conditions in the manufacturing and financial sectors. The industries mentioned are classified as primary research sectors.

To assess industry suitability, the HHI index is utilized to evaluate the market size of industries that fulfill the industrial requirements of Chaoyang District. The top 10% of foreign-funded enterprises within the primary research industry categories are identified based on the HHI index. Given the diverse range of enterprises, a more focused analysis will be conducted on foreign-funded companies in highly adaptable sectors, specifically leasing and business services, information transmission, software and IT services, and scientific research and technology services. Additionally, the market analysis for the financial and manufacturing sectors will incorporate the HHI index distribution, examining the market scale and industrial concentration by analyzing the top ten industries according to the HHI index, as illustrated in Table 4.

Table 4: HHI Index of Financial Industry (TOP10)

| **Company name** | **Current market value millions of dollars** | **Market share** | **HHI** |
|---|---|---|---|
| Berkshire Hathaway Inc. | 532605.9 | 0.063833 | 40.74705 |
| Lvmc Holdings Co,.Ltd | 483879.3 | 0.057993 | 33.63244 |
| Ishares Core S&P 500 ETF | 451160.5 | 0.054072 | 29.23791 |
| Vanguard S&P 500 ETF | 438436.1 | 0.052547 | 27.61194 |
| Vanguard Total Stock Market ETF | 387054.1 | 0.046389 | 21.51928 |
| Powershares QQQ Trust, Series 1 | 256310.3 | 0.030719 | 9.436616 |
| Vanguard Ftse Developed Markets ETF | 131624.5 | 0.015775 | 2.488614 |

The HHI index effectively indicates the relative size of companiesAccording to its formula, a monopoly situation occurs when the HHI index equals 1, while a scenario where all companies are of equal size results in X_1=X_2=...=X_n=1/n. An analysis reveals that the concentration of the financial market in the international arena is low, indicating a lack of industry monopolies. However, the top five firms based on the HHI index hold nearly the entire market share, and there is a notable difference in the HHI index between the Vanguard Total Stock Market ETF and the Powershares QQQ Trust, Series 1.

Table 5: Manufacturing HHI Index (TOP10)

| **Company name** | **Current market value millions of dollars** | **Market share** | **HHI** |
|---|---|---|---|
| PT Chandra Asri Pacific TBK | 6.66E+08 | 0.984473 | 9691.879 |
| PT SRI Rejeki Isman TBK | 2986018 | 0.004413 | 0.194744 |
| PT Citra Tubindo TBK | 1416658 | 0.002094 | 0.043834 |
| PT Tri domain Performance Materials TBK | 1247721 | 0.001844 | 0.034003 |
| PT Toba Pulp Lestari TBK | 624997.5 | 0.000924 | 0.008532 |
| PT Goodyear Indonesia TBK | 508400 | 0.000751 | 0.005645 |
| PT Polypheme Indonesia TBK | 482258.3 | 0.000713 | 0.00508 |
| PT Aluminide Light Metal Industry TBK | 270936 | 0.0004 | 0.001603 |
| Neoimmunetech INC | 176658.5 | 0.000261 | 0.000682 |
| Cintac S.A. | 151824.9 | 0.000224 | 0.000503 |

Unlike financial markets, the manufacturing sector exhibits a significant level of concentration, leading to the establishment of market monopolies. The data indicate that eight out of the top ten companies are Indonesian firms involved in petrochemicals, integrated textiles and clothing, pipeline processing, and petroleum-related manufacturing. There may be instances of regional monopolies, but upon reviewing the strategic planning for industrial clusters in this area, it appears that there is currently no transparent or defined strategy for the country. Consequently, the manufacturing industry was excluded as the primary focus of this study.

### *3.2. Spatial distance*

Using the Haversine formula, the absolute distance from each enterprise's registered address (city) to Beijing is calculated based on the chosen industry revenue level and suit ability. Enterprises closer to Beijing are assumed to have greater efficiency in terms of regional exchange and knowledge transfer. Additionally, to assess the combined effect of spatial distance and industry revenue in a cohesive manner and present the results more clearly, both scoring indicators are ranked on a scale from 1 to 10 based on their percentage scores. The scores of the two indicators were then summed and the top five enterprises were selected for further analysis.

### *3.3. Result*

#### 3.3.1. Leasing and business services

The leasing and business service industry is mainly concentrated in the business service industry. According to the HHI index, the industry market concentration is high, and Amazon is the world's top business service enterprise. In this case, the geographical location cannot offset Amazon's advantages in industry suitability and revenue, indicating that Amazon is more likely to want to explore new regions and has sufficient funds to invest in Chaoyang District.

Table 6: Aggregate score table for leasing and business service

| Company name | Current market value millions of dollars | Market share | HHI | Distance | HHI decile | Distance decile | Total decile |
|---|---|---|---|---|---|---|---|
| Amazon.Com, Inc. | 1964355 | 0.351791 | 1237.569 | 9800 | 10 | 4.84 | 14.84 |
| AIR Busan Co., Ltd. | 265.6558 | 4.76E-05 | 2.26E-05 | 1800 | 0 | 9.05 | 9.05 |
| Annil Co.,Ltd | 293.7439 | 5.26E-05 | 2.77E-05 | 1800 | 0 | 9.05 | 9.05 |
| Aiai Group Corporation | 27.45986 | 4.92E-06 | 2.42E-07 | 2100 | 0 | 8.89 | 8.89 |
| Aigan CO LTD | 25.7607 | 4.61E-06 | 2.13E-07 | 2100 | 0 | 8.89 | 8.89 |

3.3.2. Scientific research and technical services

The scoring framework for scientific research and technology services resembles that for leasing and business services. By definition, a company is considered to have monopolistic traits if its market share is 75%. Prestige Biopharma Limited is identified as a key investment target in Chaoyang District owing to its significantly higher HHI value compared to other sectors, even though it does not meet the criteria for being classified as a monopoly.

Table 7: Aggregate score table for leasing and business services

| Company name | Current market value millions of dollars | Market share | HHI | Distance | HHI decile | Distance decile | Total decile |
|---|---|---|---|---|---|---|---|
| Prestige Biopharma Limited | 486778.9 | 0.508498 | 2585.699 | 8200 | 10 | 5.9 | 15.9 |
| Melrose Industries PLC | 9757.245 | 0.010193 | 1.038889 | 1300 | 0 | 9.35 | 9.35 |
| Innocare Pharma Limited | 1053.91 | 0.001101 | 0.012121 | 1300 | 0 | 9.35 | 9.35 |
| Galapagos NV. | 1938.353 | 0.002025 | 0.041 | 1500 | 0 | 9.25 | 9.25 |
| Summit Therapeutics INC | 3727.512 | 0.003894 | 0.151619 | 1800 | 0 | 9.1 | 9.1 |

3.3.3. Information transmission, software and information technology services

Unlike leasing and business services as well as scientific research and technology services, competition among companies in the fields of information transmission, software, and information technology services is much fiercer, with various companies possessing unique strengths. Consequently, a thorough comparison and analysis of the detailed corporate information of these enterprises will be conducted, considering the specific circumstances of the industry in Chaoyang District.

An analysis of the application areas of five companies, along with the current land scape in China, reveals that the enterprise application sectors of the top four countries are quite saturated in the Chinese market and possess well-established technology. In contrast, Palo Alto Networks, Inc. focuses on technology that safeguards digital life, which is a significant concern for the Chinese population in the big data era, yet there are no credible competitors in China. Attracting investment from this company could enhance the development of data protection in Chaoyang District and serve as a basis for promoting investment or facilitating knowledge exchange in this sector.

Table 8: Summary score table for information transmission, software and information technology services

| Company name | Current market value millions of dollars | Market share | HHI | Distance | HHI decile | Distance decile | Total decile |
|---|---|---|---|---|---|---|---|
| Uber Technologies, Inc. | 146592.4 | 0.06152 | 37.84661 | 10200 | 10 | 4.63 | 14.63 |
| Adevinta ASA | 143372.3 | 0.060168 | 36.20218 | 7400 | 9.57 | 6.11 | 15.68 |
| Shopify INC | 127837.7 | 0.053649 | 28.78209 | 11000 | 7.6 | 4.21 | 11.81 |
| At&T Inc. | 122466.4 | 0.051395 | 26.41425 | 11600 | 6.98 | 3.89 | 10.87 |
| Palo Alto Networks, Inc. | 98713.51 | 0.041427 | 17.16159 | 10200 | 4.53 | 4.63 | 9.16 |

3.3.4. Scientific research and technical service

According to the above analysis, there is little difference in HHI index among the top five enterprises in the financial field, indicating that their market sizes are relatively consistent. Through enterprise retrieval, it is found that Lvmc Holdings Co,.Ltd, which ranks the first, belongs to the holding office, and its main holding company is the manufacturing and distribution of automobiles and motorcycles. Does not belong to a financial company, so choose to delete the first row of data and select Berkshire Hathaway Inc. As a target enterprise to attract investment

Table 9: Financial sector summary score table

| Company name | Current market value millions of dollars | Market share | HHI | Distance | HHI decile | Distance decile | Total decile |
|---|---|---|---|---|---|---|---|
| Berkshire Hathaway Inc. | 532605.8756 | 0.063833 | 40.7470 | 10576.04 | 10 | 4.11 | 14.11 |
| Ishares Core S&P 500 ETF | 451160.4883 | 0.054072 | 29.2379 | 10576.04 | 7.18 | 4.11 | 11.29 |
| Vanguard S&P 500 ETF | 438436.086 | 0.052547 | 27.6119 | 10576.04 | 6.78 | 4.11 | 10.89 |
| Vanguard Total Stock Market ETF | 387054.1341 | 0.046388 | 21.5192 | 10576.04 | 5.28 | 4.11 | 9.39 |
| Ishares Msci Japan ETF | 16858.8324 | 0.002020、 | 0.04082 | 2107.17 | 0.01 | 8.83 | 8.84 |

## 5. Conclusions

Using the development of a foreign investment attraction network in Chaoyang District as a case study, a recommendation framework was established to evaluate the economic environment of the district based on multidimensional proximity theory. It has been observed that Chaoyang District's industrial structure is primarily focused on the business services, technology services, and information technology sectors. Although there are a significant number of traditional industries, collaboration among them is minimal, and foreign enterprises have a limited impact on their advancement. By contrast, there is strong collaboration among companies in the technology and information sectors, with a high demand for foreign enterprises. This case demonstrates that the established framework is beneficial for local governments in identifying high-quality foreign-funded companies. Utilizing this framework allows relevant departments to expedite the enterprise selection process and identify suitable companies or targets that have not yet invested in China, based on the specific economic development conditions of different regions, thereby encouraging foreign investment.

However, this research has certain limitations, including the absence of an appropriate quantitative measure for institutional proximity within the framework. Additionally, there may be inconsistencies between the enterprise's actual core business and the data in the database, which require further refinement. These issues will be addressed in future studies.

**Author Contributions:** Conceptualization, Liang, G. and Xie, J; methodology, Xie, J; software, Xie, J and Qiu, X; validation, Liang, G and Zhang, S.; formal analysis, Xie, J; investigation, Xie, J and Qiu, X.; resources, Liang, G.; data curation, Liu, H. and Li ,M; writing—original draft preparation, Xie ,J.; writing—review and editing, Liang, G.; visualization, Xie, J.; supervision, Zhang, S.; All authors have read and agreed to the published version of the manuscript.

**Funding:** This research was support by the National Natural Science Foundation of China (Project No. 72204014 & 72304023), and the Natural Science Foundation of Beijing,China (Project No. 9232002).

**Informed Consent Statement:** Not applicable.

**Data Availability Statement:** Data will be made available on proper request from the corresponding author.

**Conflicts of Interest:** The authors declare no conflicts of interest.